\documentclass[%
12pt,
  reprint,
 superscriptaddress,
 amsmath,amssymb,
 aps,UTF8,
]{revtex4-1}

\usepackage{graphicx}% Include figure files
\usepackage{dcolumn}% Align table columns on decimal point
\usepackage{bm}% bold math
\usepackage{xcolor}
\usepackage{url}
\usepackage{float}
\usepackage{tabularx}
\usepackage{lipsum}
\usepackage{array}

\usepackage{booktabs}
\usepackage{subfigure}
\usepackage{multirow}

\begin{document}

%\preprint{APS/123-QED}

\title{ Page Entropy of Proton System in Deep-Inelastic-Scattering at Small-$x$ }%% Force line breaks with \\
%%\thanks{A footnote to the article title}%
\author{Wei Kou}
\email{kouwei@impcas.ac.cn}
\affiliation{Institute of Modern Physics, Chinese Academy of Sciences, Lanzhou 730000, China}
\affiliation{University of Chinese Academy of Sciences, Beijing 100049, China}
\author{Xiaopeng Wang}
\email{wangxiaopeng@impcas.ac.cn}
\affiliation{Institute of Modern Physics, Chinese Academy of Sciences, Lanzhou 730000, China}
\affiliation{University of Chinese Academy of Sciences, Beijing 100049, China}
\affiliation{Lanzhou University, Lanzhou 730000, China}
%\author{Chengdong Han}
%\email{chdhan@impcas.ac.cn}
%\affiliation{Institute of Modern Physics, Chinese Academy of Sciences, Lanzhou 730000, China}
%\affiliation{University of Chinese Academy of Sciences, Beijing 100049, China}
\author{Xurong Chen}
\email{xchen@impcas.ac.cn (Corresponding author)}
\affiliation{Institute of Modern Physics, Chinese Academy of Sciences, Lanzhou 730000, China}
\affiliation{University of Chinese Academy of Sciences, Beijing 100049, China}
\affiliation{Guangdong Provincial Key Laboratory of Nuclear Science, Institute of Quantum Matter, South China Normal University, Guangzhou 510006, China}


\begin{abstract}
The partons model reveals the dynamical structure of nucleons (protons and neutrons). Studies related to thermodynamic quantities of nucleons are interesting and topical questions. In this work, for the first time we apply Page's theory of the studies of black hole to investigate the entropy of proton system. Inspired by the quantum entanglement entropy in black hole information theories, we establish the proton entanglement with the similar way. Based our calculations, the proton entanglement entropy has the approximate form $S = \ln m -1/2$, where $m$ represents the partons density of proton (mainly gluon and sea quark contributions) at small Bjorken $x$. Our calculations using Page's theory are well in agreement with the recent DIS measurements from H1 Collaboration.  

\end{abstract}

\pacs{24.85.+p, 13.60.Hb, 13.85.Qk}% PACS, the Physics and Astronomy
%% \keywords{Suggested keywords}%Use showkeys class option if keyword
\maketitle

%\tableofcontents

\section{Introduction}
\label{sec:intro}
Hadrons make up the great majority of matter in the entire world. A large number of high-energy hadron collision experiments have led to more information about the internal structure of hadrons. QCD provides a powerful tool for studying strong interactions, but the annoying color confinement effect has never allowed a fascinating glimpse into the details of strong interactions in the infrared region.
There may be a deep connection between infrared behaviour of strong interaction and gravity. An excellent example is the AdS/CFT correspondence that has been developed in recent decades \cite{Maldacena:1997re}. One always separates the scales of the studied objects to accommodate the physical laws. For example, the macroscopic scale of gravity and the microscopic scale of quantum theory. Such views are scientific and rigorous, although one is more interested in thinking about microscopic properties at large scales as well as quantum properties. 

A striking recent example of a specific link between gravity and QCD is the so-called BCJ ``double copy", where the perturbative gravity amplitude consists of the QCD amplitude and an additional kinematic factor \cite{Bern:2008qj}. In a recent work, the authors have classicalized and unitarized the objects of QCD and gravity theories: the CGC-black hole correspondence \cite{Dvali:2020wqi,Dvali:2021ooc}. The remarkable link between the descriptions of black holes as highly occupied condensates of $\mathbf{N}$ weakly interacting gravitons and that of color glass condensates (CGCs) as highly occupied gluon states have been fully discussed. The authors of Ref. \cite{Dvali:2021ooc} also give the entropy of CGCs and black hole which are equal to the area measured in units of the Goldstone constant corresponding to the spontaneous breaking of Poincar$\acute{\mathrm{e}}$ symmetry by the corresponding graviton or gluon condensate. The link between black hole and collisions of nucleus is well-founded, and the above discussion provides a window to investigate the similarity between gravity and strong interaction. 

In this work, we focus our research on the proton system. Since the establishment of the partons model \cite{Bjorken:1968dy,Bjorken:1969ja,Feynman:1969ej,Feynman:1969wa,Gribov:1973jg,feynman2018photon}, it has been used in various research areas of high energy collision. In consequence, the evolution of the partons density leads to differences in the proton structure functions with different experimental inputs. The dynamical quantities in the Deep-Inelastic-Scattering (DIS) process are defined as the Bjorken scale $x$ and the photon virtuality $Q^2$. In DIS, virtual photon is used as a probe to interact with proton target to detect the proton structure, and the partons model provides good agreements with DIS data well. Experimental measurements with high statistical data also provide the important evidences for understanding the hadron structure. Both the structure of the partons density inside the proton and the multiparticle scenario of the final state of the DIS process are closely related to statistical physics \cite{Iancu:2004es,Peschanski:2006dm,Peschanski:2006dy,Munier:2009pc,Mueller:2014fba}. In these contexts, theoretical or experimental aspects of high-energy collision processes regarding entanglement entropy come into the limelight.

Recently, Kharzeev and Levin proposed the entanglement entropy as an observable in DIS \cite{Kharzeev:2017qzs}, and suggested a relation between the entanglement entropy and partons distribution. In this view, the proton in its rest frame is described by a pure quantum state which corresponds a zero von Neumann entropy. However, on one hand, in the partons model, if one does not consider the coherence between quasi-free partons, the non-zero entropy provided by the partons in proton is emerged \cite{Kharzeev:2017qzs}. On the other hand, in DIS case, one should think that the photon probe only a part of whole proton wave function denoted region $\mathbf{A}$. The DIS probes the spatial region $\mathbf{A}$ localized within a tube of radius
$\sim 1/Q$ and length $\sim 1/(m_px)$ \cite{Gribov:1965hf,Ioffe:1969kf}, where $m_p$ is the proton mass. The inclusive DIS processes sum over the unobserved final state information -- the un-probed region of proton wave function denoted region $\mathbf{B}$. Therefore only the reduced density matrix of proton $\hat{\rho}_A=\mathrm{tr}_B\hat{\rho}$ should be considered. Hence, the von Neumann entropy arising from the quantum entanglement between regions $\mathbf{A}$ and $\mathbf{B}$ \cite{Kharzeev:2017qzs,Kharzeev:2021yyf}. To deal with the entanglement entropy of protons, one can refer a lot of works which have given several helpful answers and discussions \cite{Levin:2019fvb,Tu:2019ouv,Kharzeev:2021nzh,Kharzeev:2021yyf,Levin:2021sbe,Zhang:2021hra,Hentschinski:2021aux,Hentschinski:2022rsa}. In particularly, the authors in Ref \cite{Zhang:2021hra} represented the DIS process as a local quench in Lipatov’s spin chain \cite{Lipatov:1993yb} and studied the time evolution of the produced entanglement entropy, which suggested that the DIS process can be efficiently simulated on quantum
computers.  

From the different perspective, we introduce the definition of the proton entanglement entropy inspired by the QCD-gravity theory correspondence \cite{Dvali:2021ooc}. We borrow the description of the black hole entanglement entropy to establish another one -- the proton system. In order to construct the entanglement entropy formalism, one can refer to Page's conjecture of the entanglement entropy between subsystems of quantum states \cite{Page:1993df,Page:1993wv} and consider the proton as a pure quantum state composed of two subsystems, which is similar to the black hole application by Page's work \cite{Page:1993wv}. We apply Page's entanglement entropy formula to the proton quantum system for the first time and find that the entanglement entropy of the proton is consistent with recent DIS measurements by the H1 Collaboration \cite{H1:2020zpd}. Based on our calculation, it seems to be a successful application of the treatment of black hole-related problems to the study of proton structure. The organization of the paper is as follows. The Page's conjecture and the construction of proton partons distribution are reviewed in Sec. \ref{sec:page}.  The results we obtained by using Page entropy and compared with experiments are shown in Sec. \ref{sec:dissc}. At the end, some discussions and a summaries are given in Sec. \ref{sec:summary}.

\section{Page's Curve and Proton Partons distribution functions}
\label{sec:page}

\subsection{Page's black hole entropy description}
\label{subsec:page}
\begin{figure}[H]
	\centering
	\includegraphics[width=0.5\textwidth]{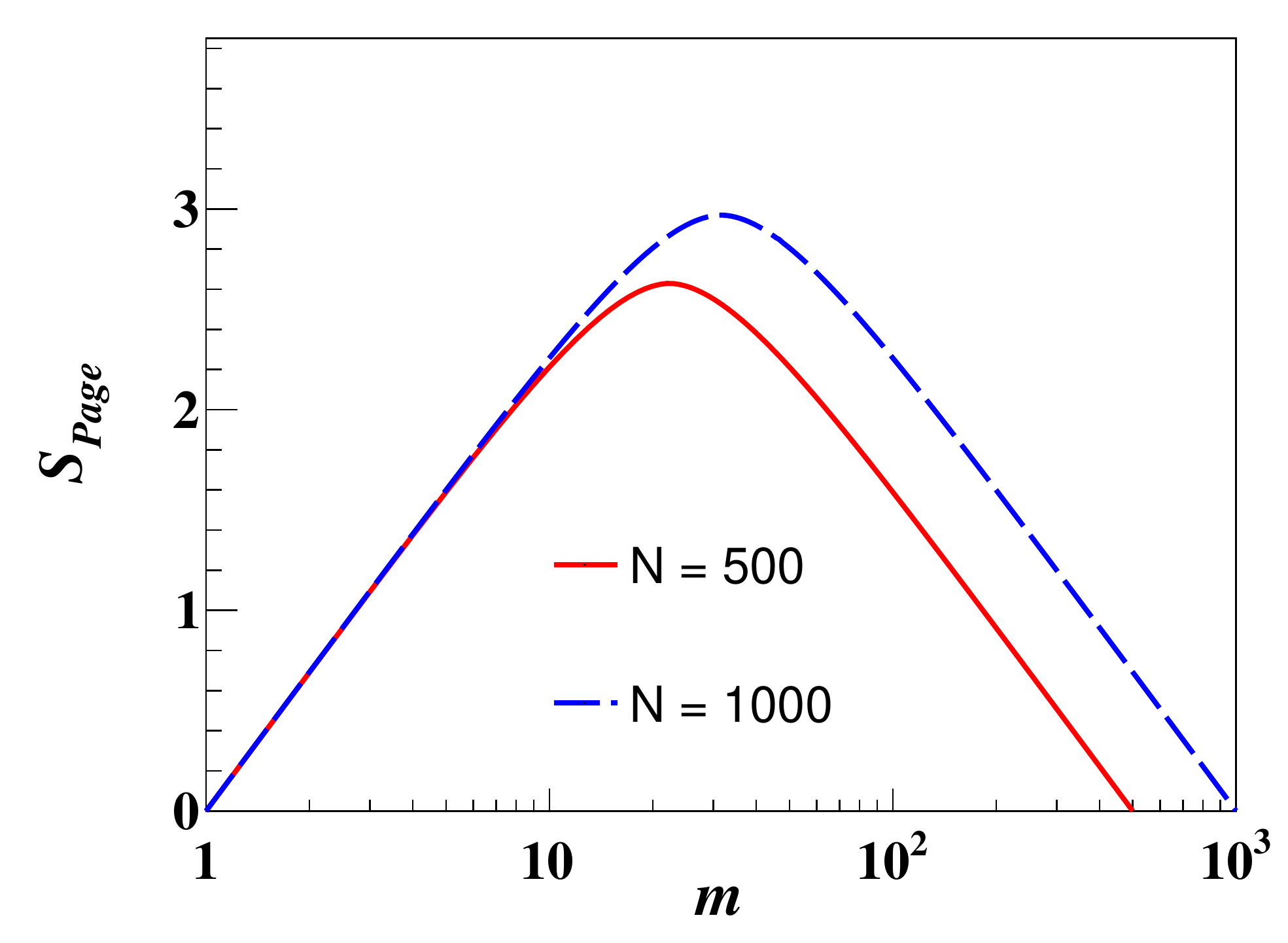}
	\caption{Example of two Page’s curves from Eq. (\ref{eq:page-entrpoy}) with total numbers of degrees of freedom 500 and 1000.}
	\label{fig:page-curve}
\end{figure}

We begin with the description of a application of Page's entanglement entropy of quantum state system. About fifty years ago, Hawking calculated and found that the classical black hole thermal radiation would lead to the final complete disappearance of the black hole, i.e. the information about the black hole would evaporate completely \cite{Hawking:1974rv,Wald:1975kc,Parker:1975jm,Hawking:1976ra}. For reviews of the problem, one can see \cite{Preskill:1992tc,Harvey:1992xk,Sanchez:1993jv,Danielsson:1993um,Giddings:1992tm} and references therein. In 1993, Page proposed that the evaporation process of a black hole can be described by $S$-matrix \cite{Page:1993df,Page:1993wv}, thus avoiding the loss of information during the process from the initial pure state black hole to the final pure state. The famous Page's curve and the Page's formula are presented from these works. We need to emphasize that Page's description of black holes originates from the entanglement entropy of quantum state subsystems. In order to obtain the entropy of a quantum system, the best way is to divide the system into subsystems and ignoring their correlations \cite{Page:1993df}. For example, consider the Hilbert space $\mathcal{H}$ of a generic quantum bipartite system, $\mathcal{H}=\mathcal{H}_A^m\otimes\mathcal{H}_B^n$, where $m$ and $n$ indicate the dimension of region $\mathbf{A}$ and $\mathbf{B}$ respectively. The total dimension is $N=mn$, and the corresponding definitions such as quantum states and density matrices can be found in \cite{Page:1993df} for details. From these concepts, one should express the average entanglement entropy of on subsystem as Page conjectured and proven by Sen \cite{Sen:1996ph},
	\begin{equation}
		S= \begin{cases}\sum_{k=n+1}^{N} \frac{1}{k}-\frac{m-1}{2 n}, \quad \text { for } m \leq n, \\ \sum_{k=m+1}^{N} \frac{1}{k}-\frac{n-1}{2 m}, \quad \text { for } m \geq n.\end{cases}
		\label{eq:page-entrpoy}
	\end{equation}
We need to emphasize that the first and second expressions in Eq. (\ref{eq:page-entrpoy}) are symmetric with respect to $m\leftrightarrow n$. If one estimates
that for $1\ll m\leq n$, Eq. (\ref{eq:page-entrpoy}) has the general and compact form \cite{Page:1993df},
\begin{equation}
	S_{m, n} \simeq \ln m-\frac{m}{2 n}.
	\label{eq:entropy-mn}
\end{equation}
FIG. \ref{fig:page-curve} shows that, for a fixed dimension of the observable subsystem, $m$, $S_A$ is bigger for bigger $N$, hence bigger $n$, the un-observable d.o.f. Page suggested that when the dimensions $m$ and $n$ of both subsystems $\mathbf{A}$ and $\mathbf{B}$ are large, and when the joint system (the black hole) is in a random pure state, the system has the maximal entropy \cite{Page:1993df}.

Based on above discussions, one can also directly make a similar treatment of the proton system in DIS process. Precisely because in DIS the proton also conforms to the bisystem characteristics \cite{Kharzeev:2017qzs}. According to Refs. \cite{Kharzeev:2017qzs,Tu:2019ouv,Kharzeev:2021yyf}, the condition $1\ll m\leq n$ is satisfied at low $x$ in DIS. In this case, the proton partons density increase with energy but has not yet reached saturation \cite{Hentschinski:2021aux}. At small $x$ all microstates of the system are equally probable and the von Neumann entropy is maximal \cite{Tu:2019ouv} because of the proton pure quantum state. Thus, Page's entropy of the proton version is constructed at low $x$ region ($x\sim 10^{-3}$). Under this condition, we consider that the mean logarithm of the multiplicity distribution of the final-state hadrons measured by DIS should correspond to the maximum entanglement entropy \cite{Kharzeev:2021yyf,Tu:2019ouv}. The Page's curve corresponds to Eq. (\ref{eq:page-entrpoy}) and FIG. \ref{fig:page-curve} which gives the information, thus one can consider that the Page's entanglement entropy becomes maximal with $m=n=\sqrt{N}$. Now Eq. (\ref{eq:entropy-mn}) reduces approximately to
\begin{equation}
	S_{m, n} \simeq \ln m-\frac{1}{2},
	\label{eq:entropy-1/2}
\end{equation}
 Eq. (\ref{eq:entropy-1/2}) is the main claim of this work which represents the entanglement entropy of the system depending on the degrees of freedom of only one of the two subsystems $\mathbf{A}$ and $\mathbf{B}$. It can be seen that the first term of Eq. (\ref{eq:entropy-1/2}) is similar to the classical Boltzmann entropy formulation, and the second term that appears on this basis comes from the entanglement entropy formula proposed by Page \cite{Page:1993df}, which includes the entanglement between  $\mathbf{A}$ and $\mathbf{B}$. Based on this claim, one can easily correspond this degree of freedom $m$ to the partons density of the proton which are described in next subsection. 

\subsection{PDFs from evolution equation}
\label{subsec:pdf}
A good deal of studies related to the dynamics of entanglement entropy have been described in detail in Refs. \cite{Kutak:2011rb,Peschanski:2012cw,Kovner:2015hga,Kovner:2018rbf,Armesto:2019mna,Duan:2020jkz,Dvali:2021ooc,Ramos:2020kyc,Hagiwara:2017uaz,Neill:2018uqw}. In order to obtain the  experimental correspondence of proton entanglement entropy, one can consider the physical significances of the two subsystems of proton. Since the maximal entanglement state satisfies that the entanglement of region $\mathbf{A}$ with $\mathbf{B}$ is equivalent to the entanglement of region $\mathbf{B}$ with $\mathbf{A}$, i.e. $S_A = S_B$ \cite{Tu:2019ouv}. The latter quantity can be reconstructed from final-state hadrons multiplicity distribution in DIS with the ``Local parton-hadron duality" \cite{Dokshitzer:1987nm} and the ``parton liberation" picture \cite{Mueller:1999fp}, which was measured by H1 Collaboration in DIS \cite{H1:2020zpd}.  For comparing with experiment, we first assume that partons density equals to the degrees of freedom of the observed region within proton through \cite{Hentschinski:2021aux}
\begin{equation}
	m=\left\langle n\left(x, Q\right)\right\rangle=x G(x, Q)+x \Sigma(x, Q),
	\label{eq:n=g+s}
\end{equation}
where $xG(x,Q)$ and $x\Sigma(x,Q)$ denote the gluon and sea quark distribution function which describe the micro-states of the observed proton region. We emphasize that $\langle n\rangle$ in Eq. (\ref{eq:n=g+s}) is from the definition in \cite{Hentschinski:2021aux} and is not the degree of freedom of the subsystem mentioned earlier. A large number of microstates are requirements for the approximation of Eq. (\ref{eq:entropy-mn}) to be valid, i.e. the small $x$ region in the DIS language. 

Let us now describe the entanglement entropy of a proton in terms of Eq. (\ref{eq:entropy-1/2}). In order to obtain the gluon distribution at the small $x$-region, there are many schemes to be selected. Both the QCD evolution equation and the partons database can give the gluon structure functions one requires. We focus on the dynamical behaviour of the gluon distribution in the small $x$ region, thus we choose the Balitsky–Kovchegov (BK) equation \cite{Balitsky:1997mk,Kovchegov:1999yj,Kovchegov:1999ua,Balitsky:2001re,Kovchegov:2012mbw} as an input to obtain the unintegrated gluon distribution $\mathcal{F}(x,k^2)$ \cite{Kutak:2004ym}
\begin{equation}
		\mathcal{F}\left(x, k^{2}\right) 
		=\frac{N_{\mathrm{c}}R_p^2}{\alpha_{\mathrm{s}} \pi}\left(1-k^{2} \frac{\mathrm{d}}{\mathrm{d} k^{2}}\right)^{2} k^{2} \mathcal{N}(k, x),
\label{eq:uninte-gluon}
\end{equation}
where $\mathcal{N}(k,x)$ encodes the forward dipole-proton scattering amplitude and is independent on impact parameter. The strong coupling constant $\alpha_{\mathrm{s}}$ is set to 0.2 \cite{Wang:2020stj}. $R_p$ denotes the proton radius which is treated as the charge radius \cite{ParticleDataGroup:2020ssz}. To reconstruct the gluon distribution, one should get the dipole-proton scattering amplitude $\mathcal{N}$, which denotes the solution to BK equation \cite{Balitsky:1997mk,Kovchegov:1999yj,Kovchegov:1999ua,Balitsky:2001re,Kovchegov:2012mbw}. We choose the previous work \cite{Wang:2020stj} on the analytical solution to the BK equation as an input to the gluon distribution, which has a good description with the vector mesons photo-productions and proton structure functions measurements \cite{Kou:2022jmg,Wang:2022jwh}. In this approach, the BK equation can be approximated as a partial differential equation--the Fisher-Kolmogorov-Petrovsky-Piscounov (FKPP) equation \cite{Munier:2003vc,Munier:2003sj,Munier:2004xu}. The analytical solution is given by homogeneous balance method \cite{Yang:2020jmt,Wang:2020stj}. In this work, the parameters of BK solution are given from \cite{Wang:2020stj} with the global fits (see \cite{Wang:2020stj} for details). In addition, the usual integrated gluon PDF can be calculated from the unintegrated gluon distribution,
\begin{equation}
	x G\left(x, Q^{2}\right)=\int^{Q^{2}} d k^{2} \mathcal{F}\left(x, k^2\right).
	\label{eq:inte-gluon}
\end{equation}

To obtain the quark PDF generated by BK or BFKL schemes, one can apply the Catani-Hautmann procedure \cite{Catani:1994sq,Hentschinski:2021aux,Hentschinski:2022rsa} but these discussions are not included in the present work. We discuss in the next section the PDFs we used to calculate the entanglement entropy, which includes the pure gluon contribution as well as the total contribution after adding the sea quark.

\section{Results and discussions}
\label{sec:dissc}
\begin{figure*}[htbp]
	\centering  %图片全局居中
	\subfigure[]{
		\label{fig:Q1}
		\includegraphics[width=0.45\textwidth]{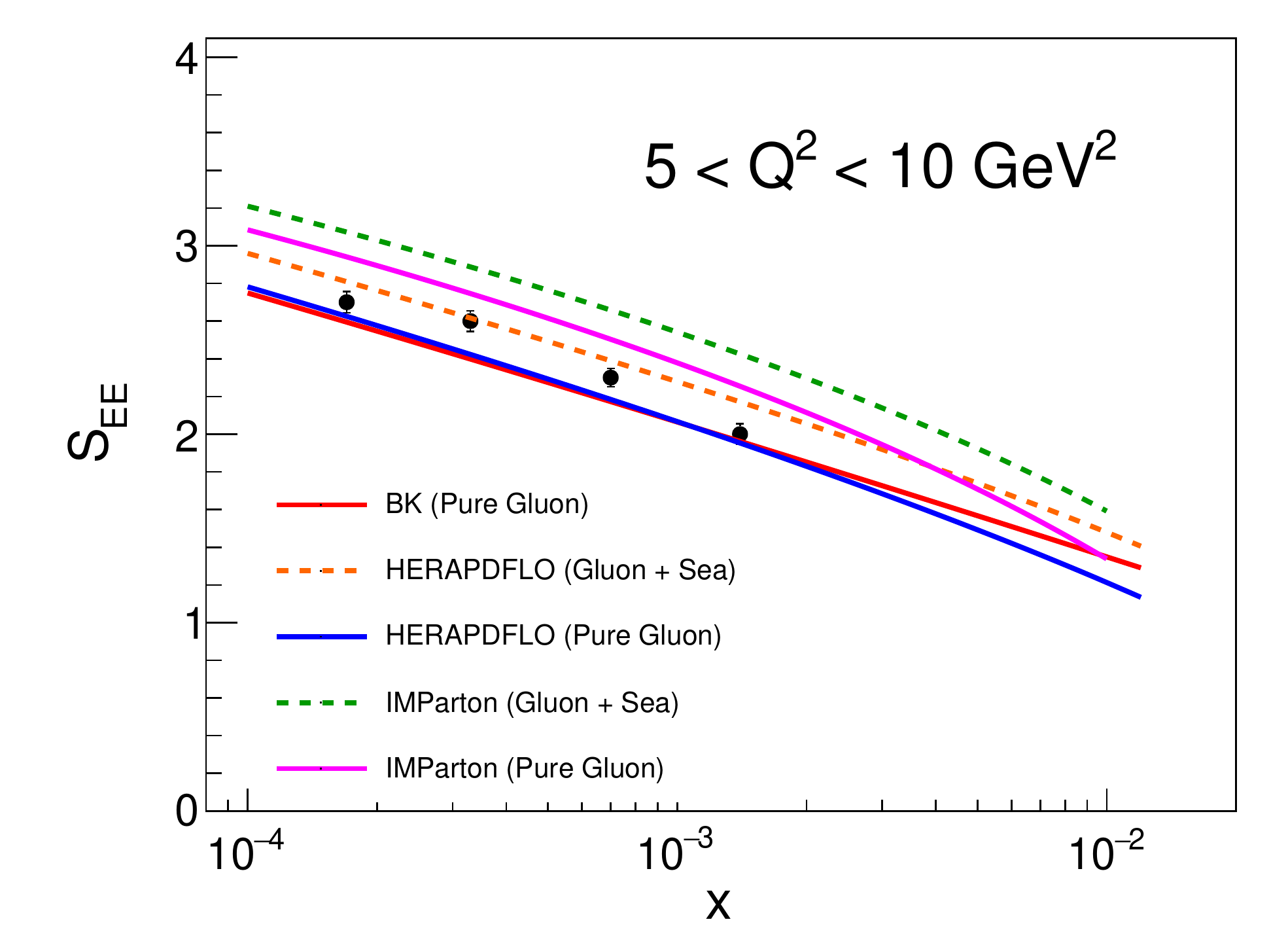}}
	\subfigure[]{
		\label{fig:Q2}
		\includegraphics[width=0.45\textwidth]{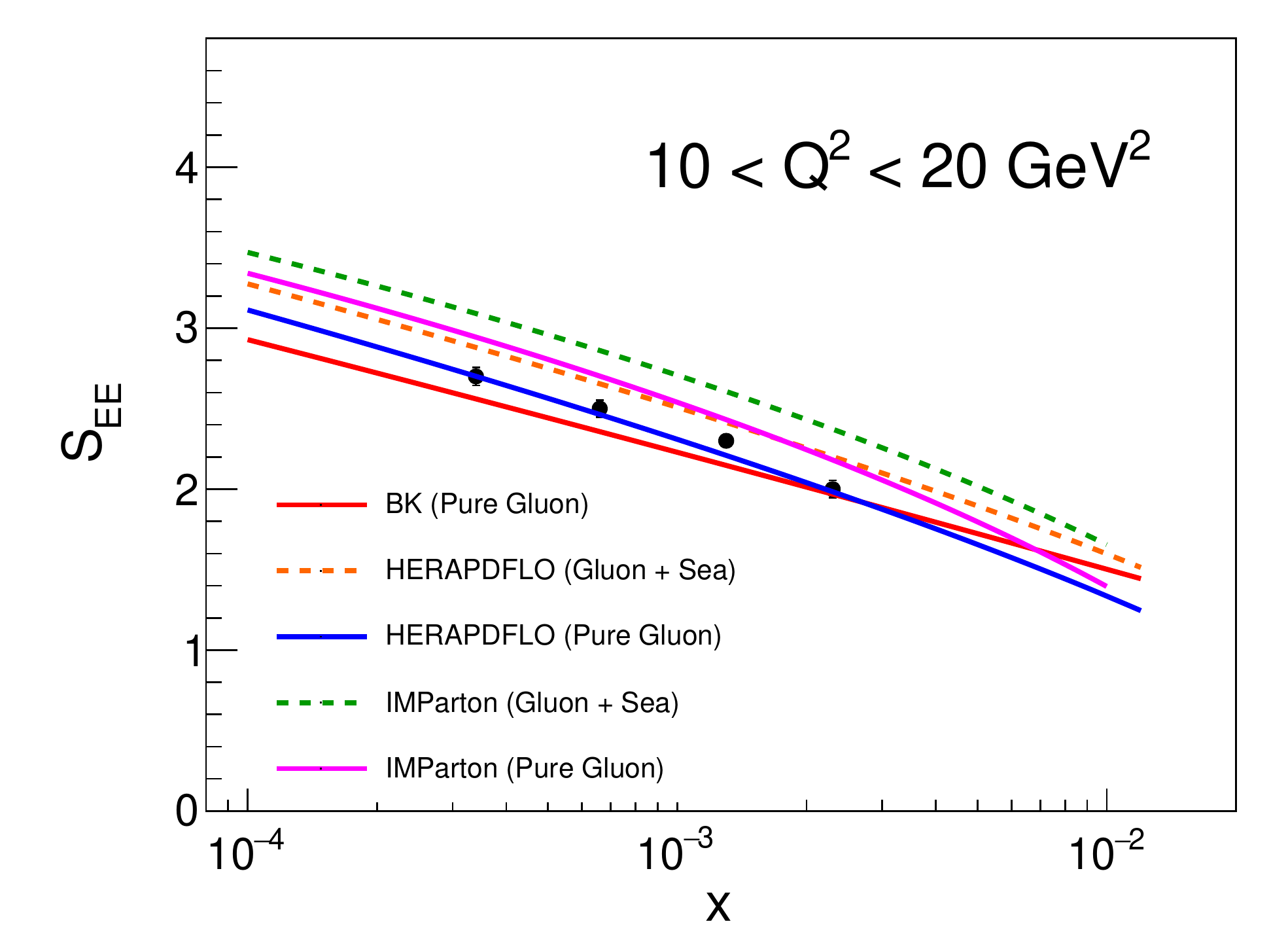}}
	\subfigure[]{
		\label{fig:Q3}
		\includegraphics[width=0.45\textwidth]{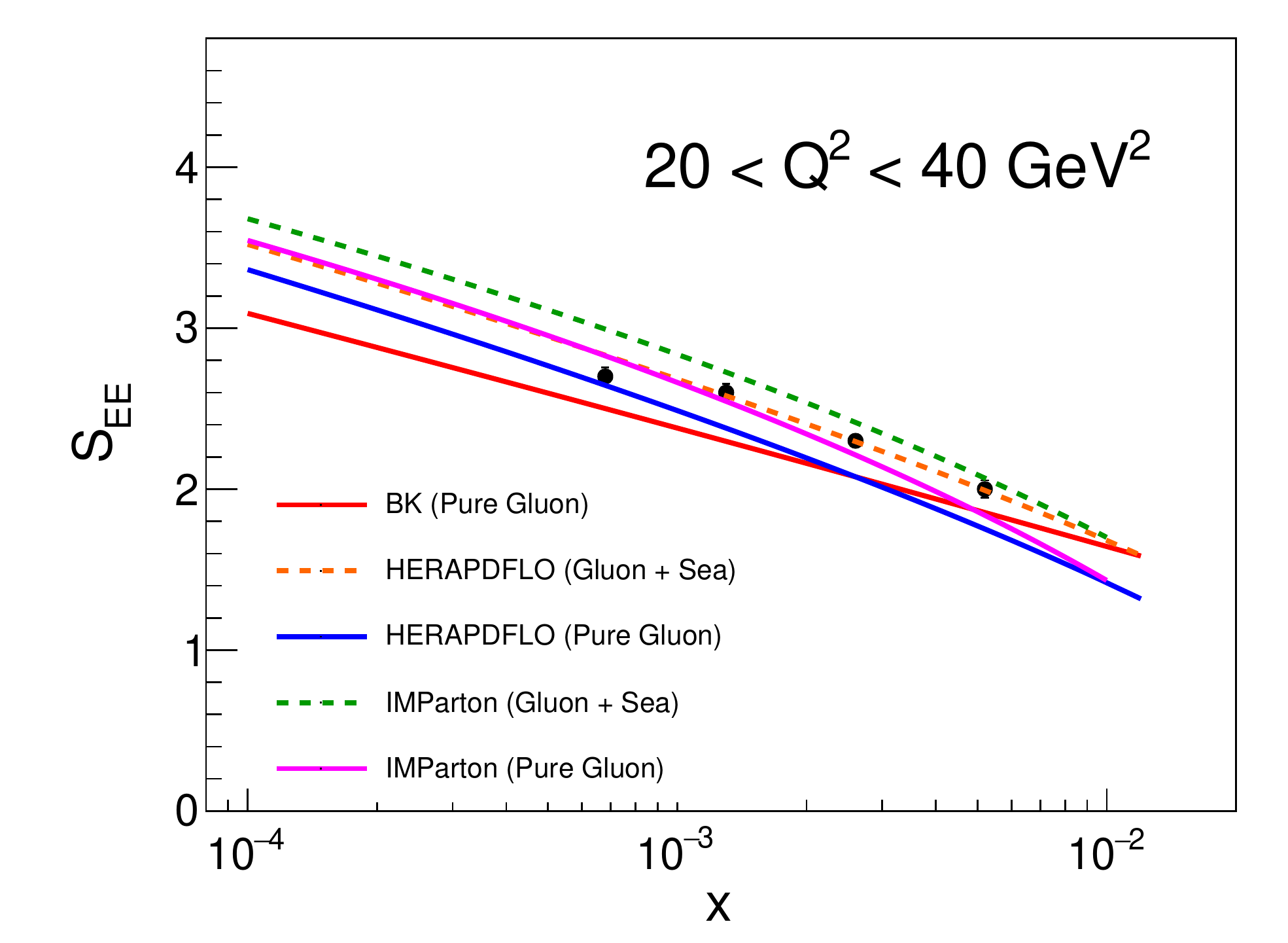}}
	\subfigure[]{
		\label{fig:Q4}
		\includegraphics[width=0.45\textwidth]{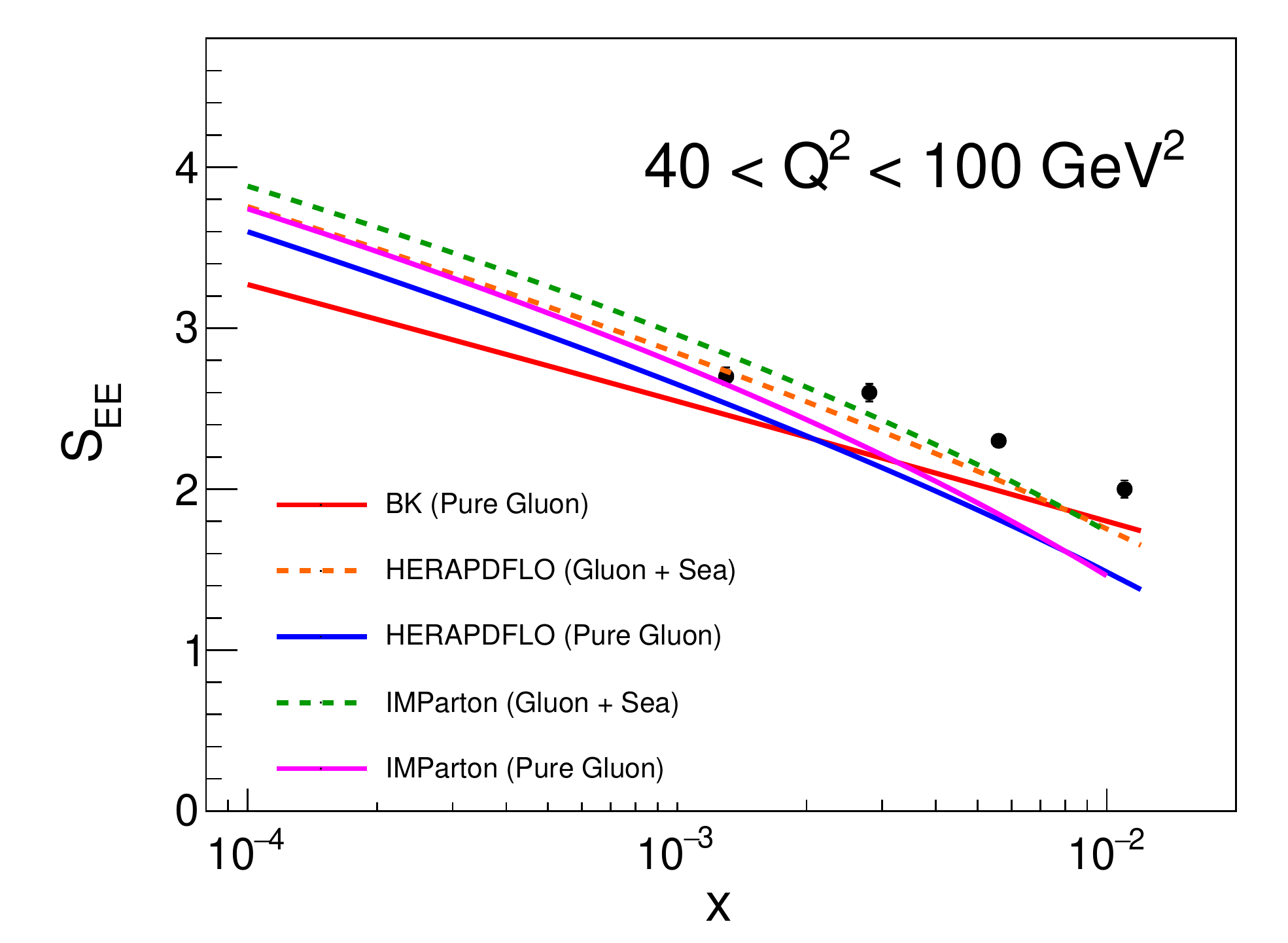}}
	\caption{Partonic entropy (colored lines) and hadronic entropy (black points) versus Bjorken $x$. The colored lines come from the BK evolution equation and partons distribution global fitting \cite{Buckley:2014ana,H1:2015ubc,Ball:2017otu,NNPDF:2014otw,Wang:2016sfq}. The solid lines represent pure gluon contribution and dashed lines represent total gluon and sea quark contribution. The results of the BK scheme (red lines) include only pure gluon contribution.}
	\label{fig:S-hardon}
\end{figure*}

To calculate entropy for the H1 $Q^2$ bins, we employ the following averaging procedure from \cite{Hentschinski:2021aux},
\begin{equation}
	\begin{aligned}
		&\bar{S}(x)_{Q_{2}^{2}, Q_{1}^{2}} \\
		&=\left(\ln \frac{\int_{Q_{1}^{2}}^{Q_{2}^{2}} d Q^{2}\left[x g\left(x, Q^{2}\right)+x \Sigma\left(x, Q^{2}\right)\right]}{Q_{2}^{2}-Q_{1}^{2}} \right)-\frac{1}{2},
	\end{aligned}
\end{equation}
where the last term comes from Eq. (\ref{eq:entropy-1/2}). Our results are shown in FIG. \ref{fig:S-hardon}. 

Based on results, we find the calculated entanglement entropies have great agreements with H1 data and the results from Ref. \cite{Hentschinski:2021aux}. We need to state that our results are all calculated under the LO approximation. It can be inferred that gluons contribution are dominant at the low $x$ region in DIS \cite{Kharzeev:2017qzs}. Sea quark only account for about $5\%$ of the total contribution. The depression behavior of the small $x$-region can also be shown in FIG. \ref{fig:S-hardon} to weaken as the scale $Q^2$ increases. Our choices for the PDFs schemes are not the focus of this work. From a formal point of view, the definition in Eq. (\ref{eq:n=g+s}) and its identification with the measured hadronic
entropy has the obvious shortcoming that it relates an unphysical object, i.e. scheme dependent parton
distribution functions (PDFs), to an observable, i.e. hadronic entropy. For example, it is well known
that the convergence of the gluon distribution is rather poor
in the low $x$ region; differences between the LO and NLO
gluon amount up to $100\%$ in the low $x$ region \cite{Hentschinski:2021aux}. However, as one of the few schemes currently linked to experimental data, we propose for the first time to connect black hole and proton entanglement entropy and partly explain the paradox that the proton entropy originally given by H1 which is relatively large \cite{H1:2020zpd}. We argue that the ``partonic" entropy should only be consistent with the entropy calculated from the measured final-state hadron multiplet distribution in DIS at the maximal entanglement state, which requires the condition of Eq. (\ref{eq:entropy-1/2}) to hold. This is the region where the extreme value of the Page's curve is located \cite{Page:1993wv} (see FIG. \ref{fig:page-curve}). 

Our idea proceeds from the quantum entangled state itself, and Eqs. (\ref{eq:page-entrpoy}-\ref{eq:entropy-1/2}) themselves are not model dependent. However, the source of the choice of gluon distribution that we use, the BK equation, and its analytical solutions are subjected to some approximations. These approximations introduce amount of uncertainties, for example, our choice of fixed strong coupling constants leads to differences in the results of gluon density for various scales \cite{Wang:2020stj}. We could not reproduce the description which is based on the collinear NNLO sea quark distribution from the work \cite{Kharzeev:2021yyf}. However, this work was one of the motivations for our research and is an inspiration for our approach. We need to emphasize that the condition for the use of Eq. (\ref{eq:entropy-mn}) is $1\ll m\leq n$, i.e. a great number of partons yields a larger subsystem degree of freedom.  

The studies and discussions related to the information entropy of black hole have made great theoretical progress in recent decades, and in Refs. \cite{Dvali:2020wqi,Dvali:2021ooc} the properties of black hole have been corresponded to the high occupied number system -- CGCs. We are inspired by this to extend the analogy of black hole systems to proton systems at small $x$. We make the assumption of formalism describing the entanglement entropy of quantum system and black hole and extend it to the measurement of the DIS process. On one hand, our first attempts seem to have been successful in terms of results. There are precedents for scaling down the investigation of the properties of macroscopic objects to the microscopic world. On the other hand, the difference between the choice of the source of the partons distribution of the proton has no effect on the main physics and the proton entanglement entropy can be constructed as long as the small $x$ condition is satisfied. The differences in the values of the same set of colored lines in FIG. \ref{fig:S-hardon} depend only on the source of the PDFs we chosen. Based on the original polynomial definition (\ref{eq:page-entrpoy}), the complete quantum pure state entanglement entropy should be determined jointly by the degrees of freedom of the two subsystems. However, as suggested by Page, it is possible to give an approximation if the subsystem degrees of freedom are sufficiently great \cite{Page:1993df}. Information on the multiplicity distribution of final-state hadrons measured by DIS as a complement to the information on the proton wave function detected by the photon probe. The latter is one of the commonly used DIS observable--structure functions or PDFs.

\section{Summary and outlook}
\label{sec:summary}
In this work, we extend for the first time the theory related to the entanglement entropy of quantum system and black hole, as formulated by Page, to describe the entanglement entropy of protons. We interpret the degrees of freedom of quantum pure state subsystems as part of the proton information probed by photons in DIS measurements, and it is this component that emerges the non-zero von Neumann entropy. We argue that the proton entanglement entropy can be expressed as $S = \ln m-1/2$ with the sum of proton gluon and sea quark distribution $m = xG+x\Sigma$, because proton can be divided into two entangled quantum systems in the DIS process \cite{Kharzeev:2017qzs}. This holds approximately in the small $x$ region where the gluon and sea quark contributions are significant. The recent DIS measurements of the H1 Collaboration \cite{H1:2020zpd} are well described by our proposal. The investigation of the partons behavior of proton at the small $x$ region will provide a more convenient tool for hardon entanglement entropy. There are in fact many relevant studies that partially explain these phenomena with small-$x$ QCD evolutions \cite{Kharzeev:2017qzs,Kharzeev:2021yyf,Kharzeev:2021nzh,Tu:2019ouv,Levin:2019fvb,Levin:2021sbe,Zhang:2021hra,Hentschinski:2021aux}. 

Last but not least, just as the BFKL and BK equation descriptions of proton gluon structure, a more exact gluon evolution at small $x$ regions may give more accurate picture of entanglement entropy, such as Jalilian-Marian-Iancu-McLerran-Weigert-Leonidov-Kovner (JIMWLK) equation \cite{Balitsky:1995ub,Jalilian-Marian:1997qno,Iancu:2000hn,Weigert:2000gi}. The direct correspondence between the gravitational theory of black hole and strong interactions is also a worthwhile subject. The entanglement entropy description of proton system is only the beginning. We argue that theoretical as well as experimental studies of CGC theory may provide a window corresponding to black hole and gravity studies.

\begin{acknowledgments}
 We would like to thank Krzysztof Kutak, Martin Hentschinski, Eugene M. Levin and M. V. T. Machado for useful correspondence. This work is supported by the Strategic Priority Research Program of Chinese Academy of Sciences under the Grant NO. XDB34030301 and Guangdong Major Project of Basic and Applied Basic Research NO. 2020B0301030008.
\end{acknowledgments}

\bibliographystyle{apsrev4-1}
\bibliography{refs}
\newpage

\end{document}